\documentclass[]{aa}  

\usepackage{graphicx}
\usepackage{txfonts}
\usepackage{lscape} 

\usepackage{xspace}

\newcommand{\Ha}{H$\alpha$\xspace}

\newcommand{\PP}{PAH$_{3.3}$\xspace}

\newcommand{\orcid}[1]{\href{https://orcid.org/#1}{\includegraphics[width=10pt]{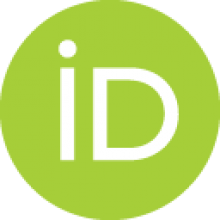}}}
\usepackage{hyperref}
\hypersetup{colorlinks, 
 linkcolor = blue,
 urlcolor = blue,
 citecolor = blue,
 anchorcolor = blue
}

\begin{document}

   \title{Not just \PP: why galaxies turn red in the Near-Infrared 
   }

 \titlerunning{Not just \PP: why galaxies turn red in the Near-Infrared}
 
   \author{Benedetta Vulcani\orcid{0000-0003-0980-1499}\inst{1}
        \and Tommaso Treu\orcid{0000-0002-8460-0390}\inst{2}
        \and Matthew Malkan\inst{2}
        \and Thomas S.-Y Lai\orcid{0000-0001-8490-6632}\inst{3}
        \and Antonello Calabr\`o\orcid{0000-0003-2536-1614}\inst{4}
        \and Marco Castellano\orcid{0000-0001-9875-8263}\inst{4}
        \and Lorenzo Napolitano\orcid{0000-0002-8951-4408}\inst{4,5}
        \and Sara Mascia\orcid{0000-0002-9572-7813}\inst{4,6}
        \and Bianca M. Poggianti\orcid{0000-0001-8751-8360}\inst{1}
        \and Paola Santini\orcid{0000-0002-9334-8705}\inst{4}
        \and Jacopo Fritz\orcid{0000-0002-7042-1965}\inst{7}
        \and Benjamin Metha  \inst{8,9}
        \and Ilsang Yoon\orcid{0000-0001-9163-0064}\inst{10}
        \and Xin Wang\orcid{0000-0002-9373-3865}
        \inst{11,12,13}
}
   \institute{INAF- Osservatorio astronomico di Padova, Vicolo Osservatorio 5, I-35122 Padova, Italy\\
              \email{benedetta.vulcani@inaf.it}
        \and
            Department of Physics and Astronomy,  University of California,  Los Angeles,  430 Portola Plaza,  Los Angeles,  CA 90095,  USA
        \and 
            IPAC, California Institute of Technology, 1200 East California Boulevard, Pasadena, CA 91125, USA
         \and 
            INAF – Osservatorio Astronomico di Roma,  via Frascati 33,  00078,  Monteporzio Catone,  Italy
          \and
            Dipartimento di Fisica, Università di Roma Sapienza, Città Universitaria di Roma - Sapienza, Piazzale Aldo Moro, 2, 00185, Roma, Italy      
        \and 
            Dipartimento di Fisica,  Università di Roma Tor Vergata, Via della Ricerca Scientifica,  1,  00133,  Roma,  Italy  
        \and
             Instituto de Radioastronomia y Astrofisica, UNAM, Campus Morelia, AP 3-72, CP 58089, Mexico          
        \and
            School of Physics, The University of Melbourne, VIC 3010, Australia
        \and
            Australian Research Council Centre of Excellence for All-Sky Astrophysics in 3-Dimensions, Melbourne, VIC 3000, Australia
         \and 
          National Radio Astronomy Observatory, 520 Edgemont Road, Charlottesville, VA 22903, USA        
         \and
         School of Astronomy and Space Science, University of Chinese Academy of Sciences (UCAS), Beijing 100049, China
         \and 
        National Astronomical Observatories, Chinese Academy of Sciences, Beijing 100101, China
        \and 
        Institute for Frontiers in Astronomy and Astrophysics, Beijing Normal University, Beijing 102206, China
}

 \authorrunning{Vulcani et al. }
   \date{}

 
  \abstract
{We measure the spectral properties of a sample of 20 galaxies at $z\sim 0.35$ selected for having surprisingly red JWST/NIRCAM F200W-F444W colors. 19 galaxies were observed with JWST/NIRSpec in the PRISM configuration, while one galaxy was observed with the high resolution gratings. 17/20 galaxies in our sample exhibit strong  3.3 $\mu m$ polycyclic aromatic hydrocarbon (PAH) emission (equivalent width EW(\PP)$\geq0.03\mu m$). In these galaxies, the strength of the color excess does not depend on environment and it  correlates with  EW(\PP). Nonetheless, the presence of the \PP alone can not fully explain the color excess, as an equivalent width of $\sim$ {  0.1}$\mu m$ is able to increase the color of galaxies by only 0.13 mag. A contribution from a hot dust component is required to explain the excess. Both the \PP equivalent width and flux correlate with the H$\alpha$ equivalent width and flux, suggesting that they are produced by the same mechanism. 5/20 galaxies showing \PP would be classified as passive based on broad band rest frame colors ((B-V) and/or UVJ diagrams) and are hence ``faux-passive". Of these, 3 galaxies have a significantly lower EW(\PP) given their color and also have low EW(H$\alpha$) and we tentatively conclude this behavior is due to the presence of an active galactic nucleus. The three galaxies with no \PP in emission have passive spectra, as do the 8 galaxies in our sample with normal F200W-F444W colors. We therefore conclude that the \PP feature is linked to dust-enshrouded star formation. The dust corrected SFR from \PP  is a factor of 3.5 higher than the SFR obtained from \Ha, suggesting that these galaxies are characterized by significant amounts of dust. }

   \keywords{Galaxies: evolution -- Galaxies: general --
                Galaxies: clusters: general 
               }

   \maketitle
%

\section{Introduction} \label{sec:intro}

Over the last decades, 
dusty galaxies have been detected by observations in the Mid- (MIR) and Far-Infrared (FIR) as well as at sub-millimeter wavelengths. Several studies have shown that they are present 
throughout cosmic history and up to a very high redshift
\citep{Smail1997, Hughes1998, 
Aussel1999, Chary2001, Blain2002, Floch2004, Marleau2004, Coppin2008, Capak2011, Reuter2023}. Their infrared luminosity mainly originates from intense dust-enshrouded bursts of star formation. But they can also host powerful active nuclei triggered by nuclear accretion onto their central black holes
\citep{Yan2005, Pope2008, Menendez2009, Hainline2009, Desai2009, Fadda2010}.

Several works have proposed that, at least at $z<1$, the amount of dust obscuration correlates with the density of the surrounding environment \citep{Duc2002, Geach2006, Marcillac2007, Bai2007, Saintonge2008, Koyama2008, Krick2009}. In the field, dust-enshrouded star formation episodes are commonly found in interacting luminous infrared galaxies \citep{Sanders1996}. In clusters,  ram-pressure \citep{Gunn1972} and harassment \citep{Moore1996} contribute to stripping the outer gas reservoir of gas–rich infalling galaxies. As a consequence, star activity is quenched in the disks while it is maintained longer in the central regions which are less influenced by the environment, but are more affected by internal extinction \citep[e.g.][]{Fritz2017}. \cite{Koyama2008} detected environmental dependence of the fraction of dusty star-forming galaxies at z$\sim$ 0.8, and 
suggested that the star formation activity detected in the MIR is enhanced in  ‘medium-density’ environments such as cluster outskirts, groups and filaments,
where galaxy optical colors sharply change from blue to red due to the truncation of star-forming activity. 
The most likely physical mechanisms at work in these medium-density regions are still debated: some works \citep{Hopkins2008, Martig2008,  Koyama2010} point to  galaxy–galaxy interaction or mergers, while others \citep{Jachym2007, Cen2014}  point to hydrodynamical effects -- which are expected to impact only the gas component, leaving the stars unaltered -- as responsible for the alteration of the star formation activity in these regions.  
However, \cite{Murata2015} did not find a clear environmental dependence of the ratio of the luminosity at 8$\mu m$ to the IR luminosity, supporting the idea that the relation between physical parameters of star-forming galaxies is not affected by the environment. 

JWST, with its unprecedented resolution and sensitivity at NIR and MIR wavelengths, is opening a new window onto dusty galaxies, enabling us to observe galaxies that were beyond the reach of previous space telescopes such as AKARI and
Spitzer \citep[e.g.][]{Evans2022, Dale2023, Belfiore2023, Lai2022, Lai2023, Lin2024, Pearson2024}.

One of the best fields studied so far with JWST is the one containing the galaxy cluster Abell 2744 (A2744) at redshift $z=0.3$ (R.A. = 00:14:20.952, decl. =-30:23:53.88). The cluster benefits from the deepest imaging available. Its central part and extended regions have 
been targeted by three different programs:  GLASS-JWST \citep{Treu2022},  UNCOVER (PID 2561, \citealt{Bezanson2024}) and the Director's Discretionary Time 
Program 2756, aimed at following up a Supernova discovered in GLASS-JWST NIRISS imaging 
(PI: Chen). Observations with a NIRCAM medium band have also been taken in Cycle 2 (GO-4111). Taken together, these surveys provide contiguous coverage in 8 bands (F090W, F115W, F150W, F200W, F277W, F356W, F410M, F444W) from 0.8 to 4.5 $\mu m$ over  46.5 square arcmin \citep{Paris2023}. 
Considering the existing 
HST imaging \citep{Lotz2016,Steinhardt2020}, the wavelength coverage is even more comprehensive. This exceptional dataset also benefits from  extended spectroscopic coverage, including JWST/NIRISS slitless spectroscopy \citep{Roberts2022}, JWST/NIRCAM slitless spectroscopy (GO-3516), JWST/NIRSpec multi-object spectroscopy  \citep{Morishita2023, Mascia2024, Price2024}, HST grism \citep{Treu2015}, and ground based spectroscopy \citep{Owers2011, Braglia2009, Mahler2018, Richard2021, Bergamini2023}. X-ray observations \citep{Owers2011} further enrich this dataset, making A2744 one of the best characterized structures beyond the local universe. 

\cite{Vulcani2023a} have characterized for the first time the near and mid-infrared colors of cluster member galaxies in A2744 and galaxies in the coeval field (0.15$<$$z$$<$0.55), unveiling the existence of a puzzling galaxy population with a surprisingly red F200W-F444W color --  called red outliers --  that does not clearly stand out in any other integrated galaxy property. 
This population is found both in  the cluster and in the field,  with a relative excess in the cluster outskirts, suggesting that environmental conditions may play a role.  
Spectrophotometric models developed to derive the properties of the stellar populations \citep{Fritz2007, Fritz2017} are not able to reproduce the observed excess in F444W, requiring additional emission from hot dust that can be simply heated by young stars.

This red excess, which is distributed across the entire galaxy disk,  resembles the IRAC excess observed by \cite{Magnelli2008} in 6 galaxies at $z>0.6$, and can have multiple explanations: (1) hot dust emission, which could originate from an obscured active galactic nucleus (AGN), (2) free-free and hydrogen recombination line emission from ionized gas, (3) a possible continuum observed in the diffuse medium of our Galaxy \citep{Flagey2006, Lu2004} and in some local star-forming galaxies \citep{Lu2003}, and  (4) emission from the 3.3 $\mu m$ polycyclic aromatic hydrocarbon (PAH$_{3.3}$) feature due to dust-enshrouded star formation. 
The latter possibility is  currently the most accredited explanation of the IRAC excess \citep{Magnelli2008, Mentuch2009}.

Prior to this publication, only one red outlier had been observed by JWST/NIRSpec, within the GLASS-JWST program \citep{Vulcani2023a}.  Its NIRSpec $3-4\mu$m spectrum, centered on the galaxy disk,  is dominated by the presence of a strong PAH$_{3.3}$ line (equivalent width of 590$\pm$ 30\AA{}, \citealt{Vulcani2023a}) that could be the source of the color excess. Only one object, however,  is not sufficient to reconstruct the general picture.

PAHs are 
found in a wide range of astrophysical sources and environments. They are prominent emission features excited by ultraviolet (UV) photons mostly from hot stars in star forming regions, while they are destroyed by hard photons from an AGN central engine \citep{Voit1992}. They can be used as a powerful diagnostic of the physical conditions in galaxies \citep{Peeters2004}. 
Their strength depends on the characteristics of the emitting particles (size, geometry, composition, charge state, etc.;
\citealt{Allamandola1989, Bauschlicher2008, Bauschlicher2009, Draine2001, Schutte1993, Tielens2008}) and  can be used to disentangle the relative contributions of starburst and AGN in galaxies \citep{Genzel1998, Rigopoulou1999, Tran2001} and as star formation tracers \citep{Peeters2004b, Tacconi2005, Tommasin2008,Kim2012, Riechers2014}.

The  PAH$_{3.3}$ feature is weak compared to the well studied PAH features observed at longer wavelengths.  
It is typically prominent in IR luminous  starburst galaxies \citep{Murakami2007, Imanishi2008, Imanishi2010, Kim2012, Lai2020} and obscured AGN associated with star formation activities \citep{Moorwood1986, Imanishi2000, Rodriguez2003}. It is attributed to the smallest PAH population which has a  small heat capacity.
Its luminosity depends on the 
radiation environment more 
strongly than the other PAH emission features, in the sense that the grains producing it are more easily destroyed by strong radiation fields \citep{Draine2007}.

Building upon the \cite{Vulcani2023a} results, in this study we further explore the origin on the red excess, and investigate more in detail the role, origin  and contribution of the PAH$_{3.3}$ feature. We present new observations of the red outliers taken with JWST/NIRSpec. In Sec. \ref{sec:data} we describe the sample selection, observations and data reduction; in Sec. \ref{sec:phot} we discuss the photometric properties of the sample, while in Sec.\ref{sec:analysis} we investigate  its spectroscopic properties, mainly focusing on the H$\alpha$ and \PP emissions. In Sec.\ref{sec:disc} we discuss the origin of the red excess and we conclude in Sec.\ref{sec:summary}.

A standard cosmology with $\Omega_{\rm m}=0.3$ $\Omega_{\Lambda}=0.7$ and H$_0$=70 km s$^{-1}$ Mpc$^{-1}$ and a \cite{Chabrier2003} Initial Mass Function are adopted.

\section{Sample selection, observations and data reduction}
\label{sec:data}

\cite{Vulcani2023a} provided a detailed characterization of the NIR properties of galaxies member of Abell 2744 and of a coeval field (0.15$<$$z$$<$0.55), exploiting the JWST/NIRCAM observations \citep{Merlin2022, Paris2023} taken as part of the GLASS-JWST survey \citep{Treu2022}. 
As mentioned in the Introduction, they are defined as ``red outliers" galaxies whose F200W-F444W color is redder than 3$\times$  width of the red sequence of the entire population. In contrast, these galaxies have typical F115W-F150W colors and rather blue rest-frame B-V colors. 
We refer to \cite{Vulcani2023a} for further details on the sample selection and results.

\begin{figure*}
    \centering
    \includegraphics[width=0.7\linewidth]{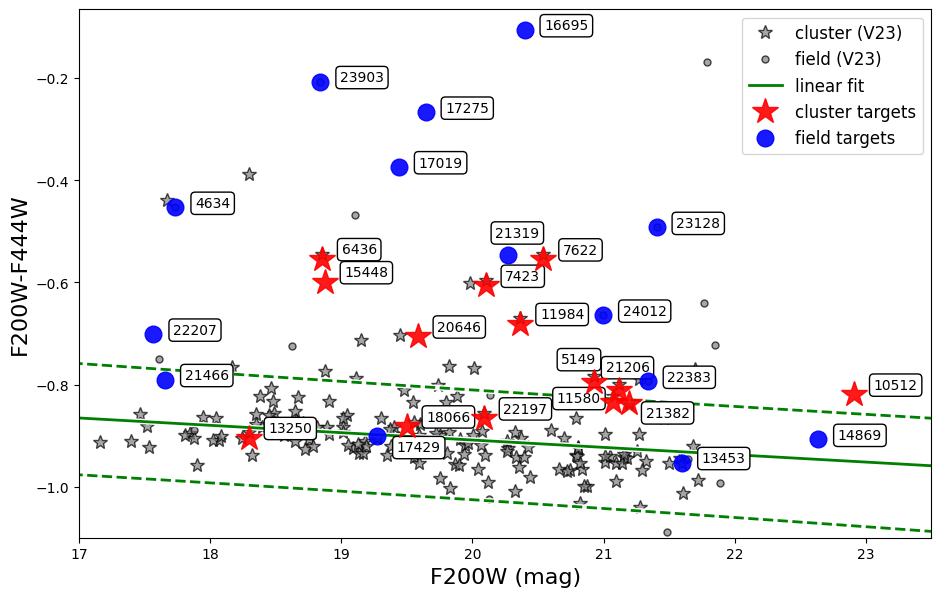}
    \caption{Observed color–magnitude diagram (F200W-F444W vs. F200W). {  Stars represent galaxies in the cluster, circles galaxies in the field. Small symbols and grey colors shows galaxies  from \citet{Vulcani2023a} not specifically targeted here. Larger symbols and red/blue colors indicate the galaxies analyzed in this paper.} 
    The green solid line represents the best fit of the relation, obtained with an iterative 3$\sigma$ clipping procedure. The dashed line shows the 3$\sigma$  error. Galaxies deviating more than 3$\sigma$  from the best fit in the F200W-F444W vs. F200W plane are defined as red outliers. Here and throughout the paper, IDs are from \citet{Paris2023}.}
    \label{fig:cmd}
\end{figure*}

The sample of cluster and field galaxies studied in \cite{Vulcani2023a} was included as fillers in two distinct JWST/NIRSpec programs for their multi-object spectroscopy observations, with a higher priority given to red outliers galaxies.

The JWST NIRSpec Program GO-3073 (PI M. Castellano) aimed at extensive follow-up spectroscopy of the $z$$\geq$10 candidates selected in the GLASS-JWST region, also observed 14 galaxies, including 11 red outliers. The same observations also included  9 galaxies as fillers in the redshift range of interest (0.15$<$$z$$<$0.55), and with spectral coverage necessary to perform the presented analysis. Some of them have the characteristics of the red outliers, while other lie on the main sequence. All of them are useful to investigate the nature of the red outliers population, so they are all included in the analysis. 
Details on observations and data reduction are given by \cite{Castellano2024} and \cite{Napolitano2024b}. The data were reduced with the STScI Calibration Pipeline version 1.13.4, and the Calibration Reference Data System (CRDS) mapping 1197. 

The JWST/NIRSpec observations taken in the context of the UNCOVER program \citep{Bezanson2024} observed 4 additional galaxies, all of them in the red outlier region. Details on observations  can be found in \cite{Price2024}. For this paper, data were taken from the catalogs published in \cite{Roberts2024} and corrected for wavelength-dependent slitloss using the NIRCam photometry and a polynomial to the residuals.

In both programs, the PRISM dispersive element was adopted, so spectra have continuous coverage in the observed wavelength range 0.6-5.3 $\mu$m, with a resolving power of $\sim$100. 

We also include in the sample the red outlier with a previously existing JWST/NIRSpec spectrum, which was already presented by \cite{Vulcani2023a}. This galaxy (41908 in \citealt{Vulcani2023a}, GLASSID 7622) was observed as part of the GLASS-JWST survey (ERS-1324, PI Treu; \citealt{Treu2022}), with JWST/NIRSpec in multiobject spectroscopy (MOS), with three spectral configurations (G140H/F100LP,  G235H/F170LP,  and G395H/F290LP),  covering a wavelength range from 1 to 5.14 $\mu$m with a spectral resolution of approximately 2700. We refer to \cite{Morishita2023,  Mascia2024} for details of the  observations and data reduction.

In total, our sample includes 28 galaxies, 20 of which are red outliers and 8 are galaxies with color within 3$\times$ the width of the red sequence identified in the F200W-F444W vs F200W plane, that we will call ``normal'' galaxies.

\begin{figure*}
    \centering
    \includegraphics[width=0.85\linewidth]{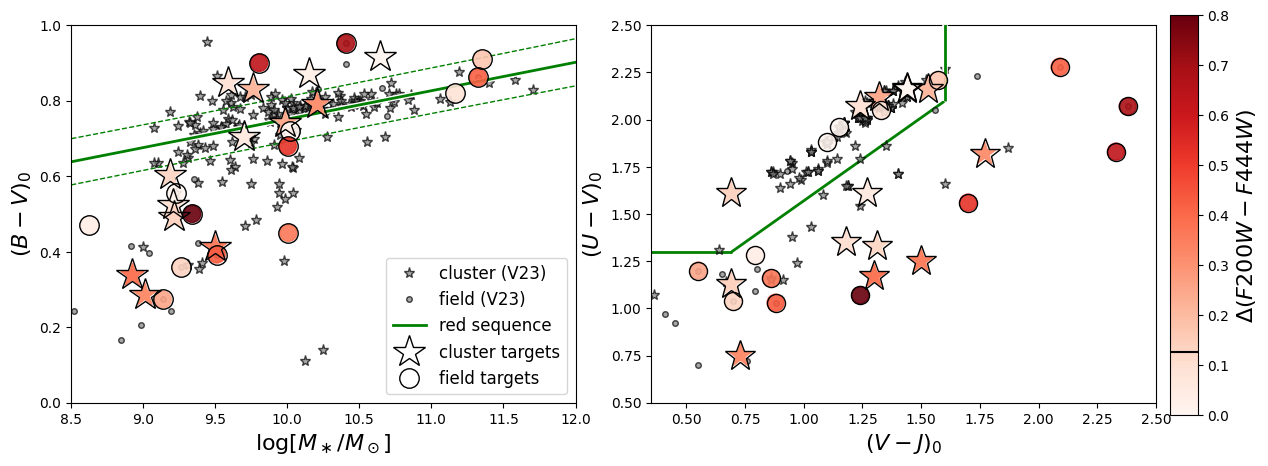}
    \caption{Rest frame colors of the galaxies. {  Stars represent galaxies in the cluster, circles galaxies in the field. Small symbols and grey colors shows galaxies  from \citet{Vulcani2023a} not specifically targeted here. Larger symbols indicate the galaxies analyzed in this paper and are color coded by their $\Delta$(F200W-F444W), which is measured as the difference between their measured color and the color they should have if they were on the red sequence, given their F200W magnitude. The horizontal black line in the color bar shows the adopted threshold between normal and red outlier galaxies.} Left: rest-frame B - V vs. stellar mass diagram. The green line represents the best fit of the red sequence, along with the 3$\sigma$ scatter. Right: rest-frame U - V vs. V - J diagram. The green line represents the separation between star forming and passive galaxies, from \cite{Labbe2005}.}
    \label{fig:rf_properties}
\end{figure*}

\section{Photometric properties of the sample of red outliers} \label{sec:phot}

Figure \ref{fig:cmd} shows the position of the 28 target galaxies on the F200W-F444W vs. F200W plane, color-coded based on the environment in which they lie.\footnote{Note that 23983,  21319, 21466, and 20646 have no F200W values. The F200W values are hence extrapolated using the F150W flux. Results are robust against this choice and stay valid if we either select galaxies in the F150W-F444W plane or exclude those galaxies from the analysis.} Observations covered most of the red outliers, and also 8 galaxies located on the red sequence. These red sequence galaxies can act as a useful comparison to understand the origin of the observed red colors. 

Even though it seems the red outliers are equally distributed between the two environments, there is a hint that in field galaxies the amount of  red excess is more significant. Given the selection function of our targets (see Sec. \ref{sec:data}), we avoid drawing solid conclusions from these statistics.

Figure~\ref{fig:rf_properties} further characterizes the spectroscopic sample in context of the entire photometric sample \citep{Vulcani2023a}, by showing the rest-frame B-V color as a function of stellar mass and the UVJ diagram \citep{Labbe2005, Williams2009}. As described in detail by \cite{Vulcani2023a}, we obtained rest-frame colors, and stellar masses, by fitting synthetic stellar templates to the available NIRCam photometry with ZPHOT \citep{Fontana2000}, following the same procedure described by \cite{Santini2023}. Focusing on the color-mass diagram, three red outliers lie above the red sequence (17275, 23903 and 20646), defined by fitting the entire cluster relation adopting a 3$\sigma$ clipping method, four are on the red sequence (22207, 4634, 15448, 11984) and the rest are in the blue cloud. 23903 is  the only galaxy in the sample that has broad lines indicative of AGNs in the MUSE spectra \citep{Vulcani2023a}. Considering the non red outliers, three are above (11580, 18066, 13250), three are on (22197, 17429, 21466) and two are below (14869, 13453) the red sequence. 
Considering the UVJ diagram, four red outliers (5149, 11984, 22207, 20646) and six normal galaxies (17429, 13453, 11580, 21466, 13250, 18066) are located in the passive region, while the rest are in the star forming/dusty region. While the two diagnostic tools, which have a different sensibility to the presence of dust, present a coherent view for the non red outliers, they depict a different picture for the red outliers, highlighting the peculiarity of the population. 

\cite{Vulcani2023a} underlined that almost half of the red outliers   
have a disturbed morphology, with field galaxies showing disturbances suggestive of ongoing/past interactions, and cluster galaxies   showing tentative signs of gas stripping, i.e. they have one-sided tails and no clear companion able to induce tidal interactions. 
Figure \ref{fig:images} shows  the color-composite images for the cluster and field targets. Half --  and actually the reddest ones --  of the cluster red outliers  show morphology consistent with ram pressure stripping (7622, 7423, 6436) and two of them have already been the subject of extended analysis at optical wavelengths that indeed confirm ram pressure stripping in action \citep{Moretti2022, Vulcani2024, Khoram2024}. Field galaxies instead show rather regular morphology, except for a few cases in which galaxies have close companions and might be undergoing some tidal effects (16695, 17019). In contrast,  galaxies 13250 and 13453 have a close companion but present normal colors. Red colors are especially evident in the ram pressure stripped galaxies and in the edge-on field galaxies. None of the images reveal the presence of clear dust lanes.

\section{Spectral analysis and Results} \label{sec:analysis}

\subsection{Spectral features}
\begin{figure*}
    \centering
    \includegraphics[width=0.7\linewidth]{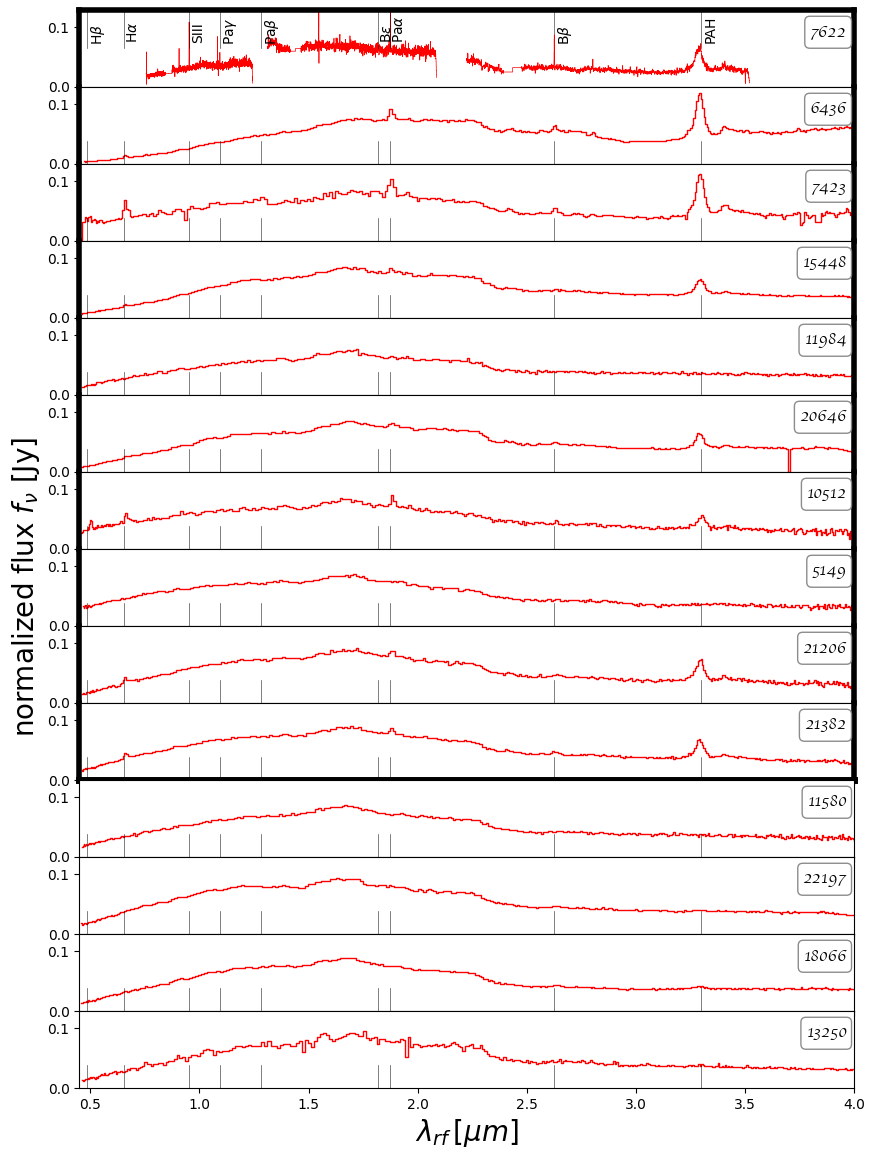}
    \caption{JWST/NIRSpec rest-frame spectra for the cluster galaxies in our sample, sorted by decreasing red excess. Highlighted within the black box are the red outliers. {  Spectra are normalized so that the flux in the rest frame range $2.9<\lambda[\mu m]<3.1$ is equal to one.}}
    \label{fig:spectra_cl}
\end{figure*}

\begin{figure*}
    \centering
    \includegraphics[width=0.7\linewidth]{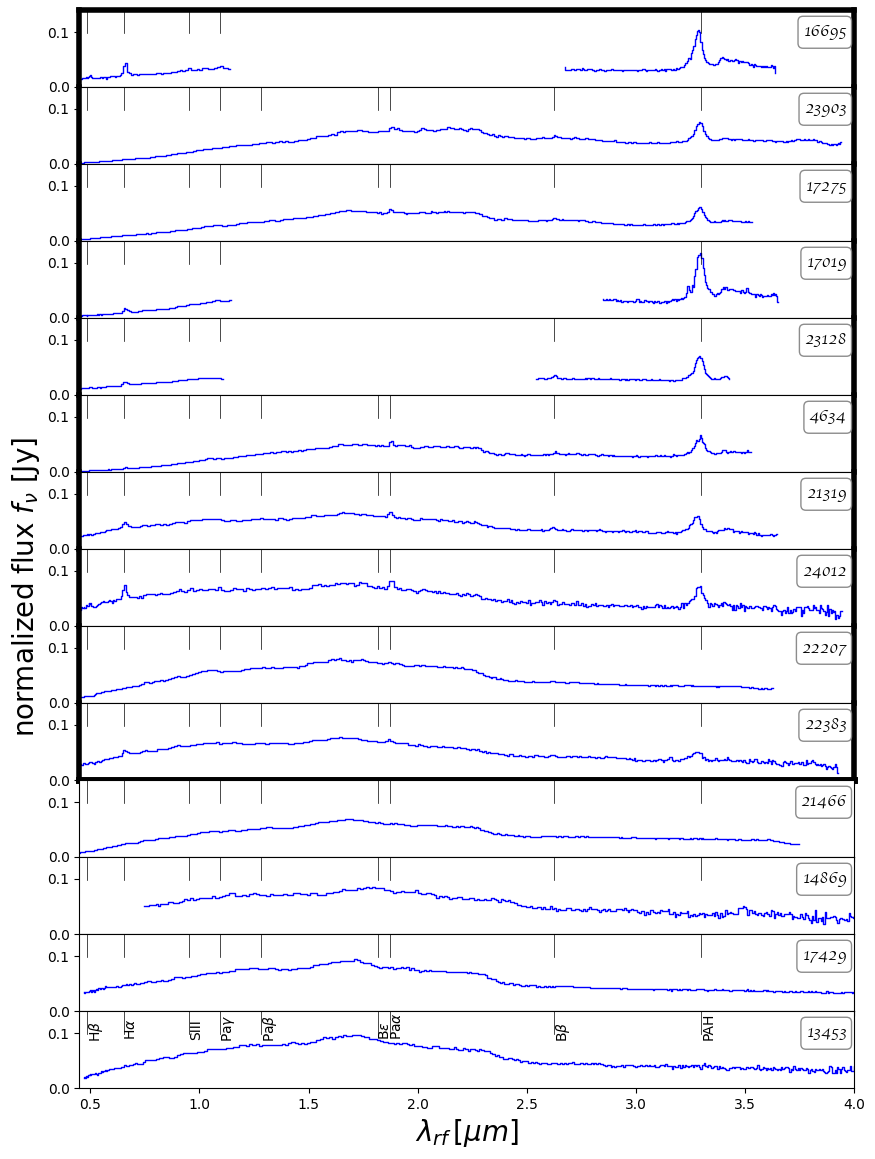}
    \caption{JWST/NIRSpec rest-frame spectra for the field galaxies in our sample, sorted by decreasing red excess. Highlighted within the black box are the red outliers. {  Spectra are normalized so that the  flux in the rest frame range $2.9<\lambda[\mu m]<3.1$ is equal to one.}}
    \label{fig:spectra_fie}
\end{figure*}

\begin{figure}
    \centering
    \includegraphics[width=0.8\linewidth]{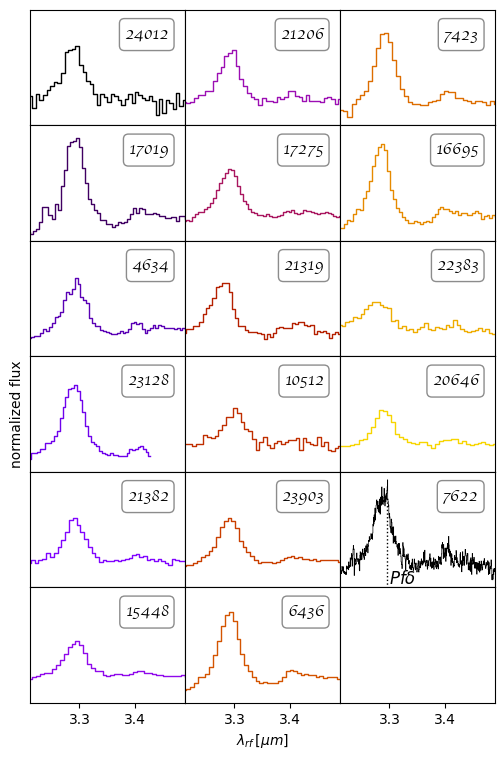}
    \caption{JWST/NIRSpec rest-frame spectra for the galaxies in our sample, zoomed-in on the PAH feature. Spectra are normalized so that the shown area is equal to one.}
    \label{fig:PAHspectra}
\end{figure}

Figures \ref{fig:spectra_cl} and  \ref{fig:spectra_fie} show the NIRSpec rest-frame spectra for all cluster and field galaxies in our sample, sorted by decreasing color excess, which is obtained as 
difference between the galaxy color and the color it would have on the red sequence given its F200W magnitude. All red outliers spectra but three (11984, 5149  and  22207) are dominated by a strong feature at 3.3$\mu$m, which is one of the PAHs. The three galaxies in which this feature is  lacking are either on the red sequence or in the UVJ passive region in Fig.~\ref{fig:rf_properties} and have passive spectra (see below) . It is interesting to note, though, that not all red sequence/UVJ passive red outliers lack the PAH feature. All the spectra of the three galaxies above the red sequence (17275, 23903, 20646) have a clear PAH$_{3.3}$ feature, but no significant other emission lines (e.g. lines of the Balmer, Paschen and Bracket series); the spectra of the remaining two galaxies on the red sequence (4634 and 15448) are characterized by both PAH$_{3.3}$ in emission as well as other emission lines. One passive galaxy according to the UVJ diagram (20646) has clear \PP in emission. There are hence 5 galaxies that one would consider passive based on broad band color, but show evidence of activity, and can be called ``faux-passive''.  In contrast, normal (i.e. not red outlier) galaxies have no significant PAH$_{3.3}$  emission.  We will return to this subject and  quantify the emission line measurements in the next section. 

For the galaxies whose spectrum shows a strong PAH$_{3.3}$ feature, we show in Fig.\ref{fig:PAHspectra} a zoom-in on the rest frame spectral region $3.2<\lambda/\mu m< 3.6$. Spectra are all normalized so that the area under the portion of the plotted spectra is equal to one. In this wavelength range, the spectrum is dominated by the PAH$_{3.3}$ band and 3.4 $\mu \text{m}$ aliphatic feature. In some cases, the two bands sit on top of a broad plateau that can extend up to  $\sim$ 3.8 $\mu m$ \citep{Allamandola1989, Joblin1995}.

There is a variety of \PP profiles.  At least in nearby sources and in our Milky Way, the peak position and width of this band have been found to be sensitive to the emission temperature (i.e. absorbed UV photon), charge and size of the emitting molecules, with positive charges and smaller molecules shifting the peak to redder wavelengths \citep{Joblin1995, Peeters2002, Diedenhoven2004}. A detailed characterization of the chemical composition and modes of the \PP is beyond the scope of this work, but it is interesting to stress the variety of profiles in our sample, indicative of a range of ongoing processes and elements. 

Two features are worth stressing: 
first of all,  the galaxy with the broadest and shallowest feature is 23903, the one most likely hosting an AGN, even though no X-ray detection is associated to this source in the archival data. This is in agreement with previous results, showing that if the equivalent width of the \PP emission is substantially smaller than that of starburst-dominated galaxies, then a significant contribution from an AGN to the observed 3–4 $\mu m$ continuum or that those small grains are destroyed by the AGN are the most natural explanations \citep{Imanishi2000, Imanishi2001, Imanishi2003, Imanishi2006b, Imanishi2006}.

Secondly, the only high resolution spectrum we have at our disposal shows evidence for the Pf$\delta$ emission, on top of the \PP emission. In the same spectrum, sub-peaks can be also be detected in the 3.4$\mu$m band, and these are consistent with the existence of different C-H stretch emitting particles \citep{Geballe1994}.

The relative importance of the 3.4 $\mu$m aliphatic feature and the 3.3 $\mu$m  aromatic feature changes from galaxy to galaxy. The variation of the 3.4 $\mu$m/3.3 $\mu$m  ratio is indicative of the processing of dust particles in the ISM. The band intensity ratio decreases for positions closer to the star \citep{Geballe1989, Joblin1995}.
In the next section we will quantify the ratio between the two.

\subsection{Emission line measurements}

22 out 28 galaxies have a previously measured spectroscopic redshift from the literature \citep{Vulcani2023a}. We determine the NIRSpec redshift of each spectrum through emission line measurements, by comparing the observed emission lines with their corresponding theoretical vacuum transitions. We give priority to the H$\alpha$ line, followed by the Pa$\alpha$ line. For galaxies with no emission lines, we adopt the previously determined redshift. In all cases, the agreement between the measured and literature redshift is of the order of 1\%. For the remaining 6 galaxies, we use the redshifts measured from the spectra.

We then measure the emission line fluxes and equivalent width (EW) for all the visible lines using the python package \textsc{LiMe} \citep{Fernandez2024}. 
We consider as reliable only lines measured with a minimum signal-to-noise ratio (SNR) of 3,  assuming a Gaussian profile. 

At the resolution of the PRISM spectra,  H$\alpha$ and [NII] cannot be deblended, so we fit them together. 

To correct for the contamination by the [NII] line doublet,  we apply the locally calibrated correction factor given by \cite{James2005}: H$\alpha$/(H$\alpha$ + [N II]) = 0.823. Given that all the galaxies but one are star forming based on the BPT diagram \citep{Vulcani2023a}, we assume this approach is appropriate.

\begin{figure}
    \centering
    \includegraphics[width=0.85\linewidth]{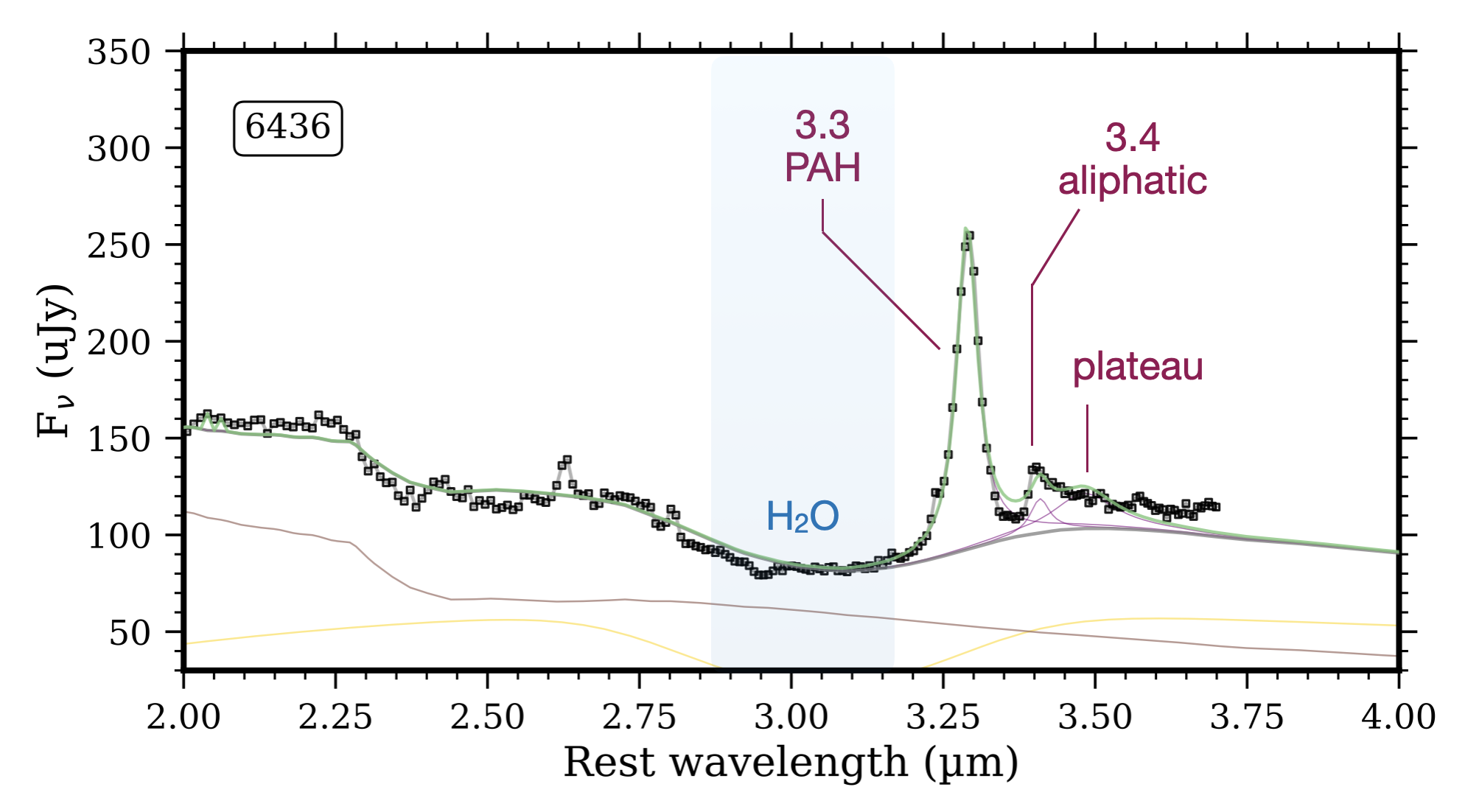}
    \caption{An example of our spectral decomposition result using CAFE on source 6436, which has strong PAH emission together with relatively high ice opacity among the sample. Features of interest are labeled, including the 3.3$\mu$m PAH feature, the 3.4 aliphatic feature, the broad plateau feature, and the water ice absorption at 3.05$\mu$m. Yellow and red lines represent stellar and non-stellar continuum sources. Note that the adopted  spectral coverage and simplified treatment do not allow us to obtain a physical significance to individual continuum components.
}
    \label{fig:PAH33_CAFEfit}
\end{figure}

We fit the NIRSpec 3$\mu$m regime using CAFE\footnote{https://github.com/GOALS-survey/CAFE}, a spectral decomposition tool that simultaneously fits the PAH features, dust continuum, and various atomic and molecular gas emission lines. This fit was conducted systematically throughout our sample. We focused on the 2.0 to 3.7$\mu$m range that includes the 3.3$\mu$m PAH feature, the 3.4 aliphatic feature,  the broad plateau feature that resides at the redder end of the PAH feature, and the water ice absorption centered at 3.05$\mu$m (as shown in Fig.\ref{fig:PAH33_CAFEfit}).

Table \ref{tab:galaxies} in Appendix A reports the extinction corrected measurements of the most important lines.

We confirm that the normal galaxies and three red outliers (11984, 5149, 22207) have no detected emission lines (S/N of the H$\alpha$ line $<5$) and we consider them to be passive. We remind the reader that the three red outliers are also passive according to the UVJ diagram (see Sec.\ref{sec:phot}). While two of them are rather close to the adopted separation between normal and red outlier galaxies, 11984 is rather red.   In the following analysis, we will exclude all the non emission line galaxies. We will discuss the origin of the red excess  in these excluded galaxies in Sec.\ref{sec:pass}. 

The ratio of the 3.4 to 3.3 $\mu m$ fluxes ranges from 0 to 0.25 (see Tab.\ref{tab:galaxies}, indicating that the 3.4 $\mu m$ aliphatic feature is relatively weak in our sample. Similar results have been found by \cite{Lai2020}. In contrast, the ratio of the 3.45 to 3.3 $\mu m$ fluxes ranges from 0.15 to 0.5, suggesting that the 3.45 $\mu m$ might give a larger contribution to the total flux. 
The origin of this feature is still under debate, and it is not clear whether an aliphatic or aromatic feature contributes more to this broad plateau feature.
In both cases, no correlation with the \PP EW  nor the color excess emerges (plot not shown).

\begin{figure}
    \centering
    \includegraphics[width=0.8\linewidth]{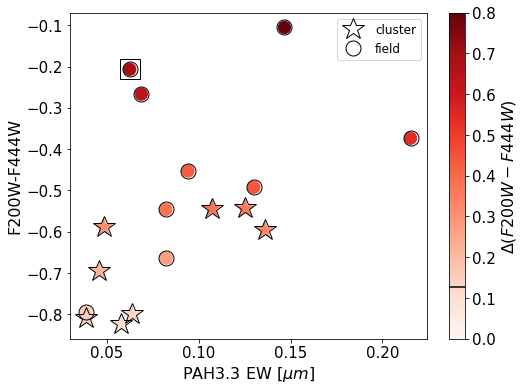}
    \caption{F200W-F444W as a function of the rest frame equivalent width of the PAH, for cluster (stars) and field (circle) galaxies. Galaxies are color coded by their $\Delta$(F200W-F444W), which is measured as the difference between their measured color and the color they should have if they were on the red sequence, given their F200W magnitude. The squared symbol indicates 23903, the only galaxy most likely hosting an AGN.}
    \label{fig:color_EW}
\end{figure}

Figure \ref{fig:color_EW} correlates the rest frame equivalent width (EW) of the PAH$_{3.3}$ with the galaxy F200W-F444W color. A  correlation is seen, with redder galaxies having the tendency of also having  higher EW values. A Spearman correlation test confirm this results, with a correlation coefficient of 0.62 and a significance of  p=0.008.  Most likely due to the small sample, no clear differences emerge between cluster and field galaxies. Symbols are color coded by the red excess. Galaxies with higher excess tend to have higher EW values. One of the  galaxies clearly stands out for having a lower EW than expected from its color is hosting the AGN, and the \PP emission might be altered by its presence. No information about the presence of AGN are available for the other two galaxies (17275 and 20646).

\begin{figure*}
    \centering
    \includegraphics[width=0.8\linewidth]{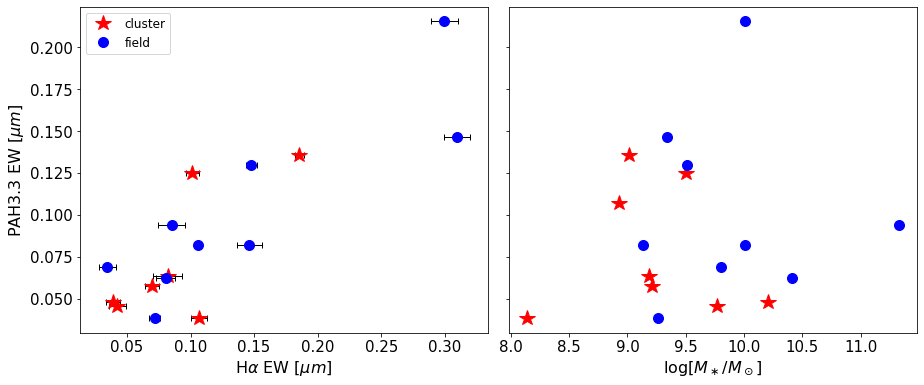}
    \caption{Rest frame equivalent width of the PAH as a function of rest frame equivalent width of H$\alpha$ (left) and stellar mass (right), for cluster (red) and field (blue) star forming galaxies. }
    \label{fig:EW_mass}
\end{figure*}

We further compare the \PP EW with the H$\alpha$ EW (for the galaxies for which this measurement is available) and with the galaxy stellar mass in Fig.\ref{fig:EW_mass}. Overall, the strengths of the EWs correlate, even though the increase in the EW of \PP is faster than the increase in the EW of H$\alpha$. A Spearman correlation test supports the correlation, with a coefficient of correlation of 0.6 and a significance p = 0.001. The EW of \PP does not correlate with stellar mass (p = 0.88), even though it seems that only low mass galaxies  might have high values of EW. Interestingly, the low mass cluster galaxies with very high EW values (7622, 22043, 18708) are all ram pressure stripped galaxies \citep{Moretti2022, Vulcani2024}, suggesting a connection between ram pressure stripping and strength of the \PP. Indeed, when RPS is at play, the compression of the gas by the ISM induces an enhancement in star formation \cite{Vulcani2018_L, Vulcani2020b}, which can appear as an increase in the \PP EW.

\begin{figure*}
    \centering
    \includegraphics[width=0.8\linewidth]{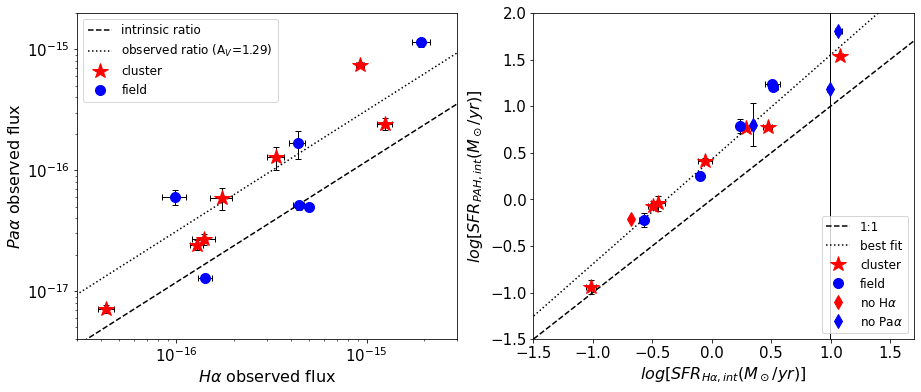}
    \caption{Left panel: Observed Pa$\alpha$ vs. H$\alpha$ fluxes, for the subset of galaxies with both measurements. The dashed line shows the intrinsic ratio, corresponding to a temperature T = 10$^4$ K and an electron density $n_e = 10^2 cm^{-3}$ for Case B recombination (Osterbrock 1989). The dotted lines corresponds to the ratio obtained assuming an E(B-V) =0.41, which is the best fit to our data. Right panel: comparison between the dust corrected SFR obtained from H$\alpha$ and \PP. In both panels,  cluster galaxies are shown in red, field  galaxies are shown in blue. }
    \label{fig:SFR}
\end{figure*}

\subsection{SFR from PAH}\label{sec:SFR}
Finally, we measure SFRs from both the H$\alpha$ and \PP fluxes and compare their values in Fig.\ref{fig:SFR}. We obtain luminosities from fluxes, taking into account the redshift of the sources. We then correct the H$\alpha$ luminosity for the presence of dust using the Pa$\alpha$ luminosity. The intrinsic H$\alpha$/Pa$\alpha$ decrement remains roughly constant for typical gas conditions in star-forming galaxies \citep{Osterbrock1989}. In our analysis, we assume the intrinsic value of (H$\alpha$/Pa$\alpha$)$_{\rm int}$ = 8.46, corresponding to a temperature T = 10$^4$ K and an electron density $n_e = 10^2 cm^{-3}$ for Case B recombination \citep{Osterbrock1989}. This choice is standard for star-forming galaxies in the literature.  We than assume a \cite{Cardelli1989} reddening curve and obtain the extinction at the H$\alpha$ wavelength. The left panel of Fig.\ref{fig:SFR} shows the comparison between the observed H$\alpha$ and Pa$\alpha$, for the galaxies with both measurements. Galaxies are characterized by a wide range of extinction (see Tab.\ref{tab:galaxies}), with a maximum A$_V$ of 3.1 (6436). The observed flux ratio is consistent with a mean A$_V$= 1.29.

We measure H$\alpha$-based star formation rates from \cite{Kennicutt1998b} for a \cite{Chabrier2003} IMF:  $SFR = 4.6\times 10^{-42}\times L(H\alpha_{int})$, with L(H$\alpha_{int}$) dust corrected luminosity measured in erg/s.

To obtain SFRs from  the \PP feature, we use the intrinsic \PP fluxes obtained by CAFE\footnote{Using the observed fluxes and correcting them using a \cite{Cardelli1989} law give consistent results.} and then use the relation  found in \cite{Lai2020}, converted to a \cite{Chabrier2003} IMF: 
$\log SFR =  -(6.80 \pm 0.18) + \log[L(PAH_{ 3.3})/L_\odot$]-0.05.
This relation has been calibrated against the  [Ne II] {  (at 12.8 $\mu m$)} and [ Ne III] {  (at 15.6 $\mu m$)} luminosities, which in turn have been calibrated by \cite{Ho2007} against the Br$\alpha$. The timescales for star formation are hence similar to those of the H$\alpha$, $<10^7$ yr.

The right panel of Figure~\ref{fig:SFR} shows a tight correlation between  the SFRs obtained from H$\alpha$ and those obtained from the \PP. Nonetheless, the values obtained from the \PP are systematically larger, by a factor of 3.5, on average. In Sec. \ref{sec:disc_SFR} we will discuss the possible origin and meaning of the discrepancy.

We note that for one galaxy (7622) we do not have the  H$\alpha$ line. We hence use the Pa$\alpha$ luminosity,  corrected for dust assuming A$_V$= 1.29 found using the other galaxies. On the contrary, for three galaxies (16695, 17019, 23128) we do not have the Pa$\alpha$ line to correct for dust the SFR form H$\alpha$. The reported value in the plot is hence corrected for dust assuming the average A$_V$= 1.29.

Finally, the 3.4/3.3 ratio seems to decrease with increasing values of SFRs (plot not shown). This is in line with previous results: \cite{Yamagishi2012} found that the ratio decreases toward the center of the starburst galaxy M82, suggesting that the relative abundance of the aliphatic band carriers drops toward regions with more intense star formation. The decline of the 3.4/3.3 ratio in photodissociation regions is consistent with this picture, suggesting that an efficient photochemical process leads to the destruction of the aliphatic subgroups present on the periphery of PAHs \citep{Joblin1996, Pilleri2015}.

\section{Discussion}\label{sec:disc}

\subsection{Explaining the red excess}\label{sec:disc}

In the previous sections we have characterized the spectral properties of the red outliers, mainly focusing on the \PP emission feature. Overall, we have identified three populations of red outliers: i) 14 emission line galaxies with also a clear \PP in emission, which are mostly located below the optical red sequence; ii) three galaxies with \PP in emission, but rather weak emission lines (S/N of the H$\alpha$ line between 5 and 10), mostly located above the optical red sequence; iii) three passive galaxies.  
As a comparison, all the non red outliers, regardless of their position on the rest frame (B-V)-stellar mass plane,   are passive with no \PP in emission. 

In this section we will connect these results, proposing a scenario able to explain our findings.

\subsubsection{Emission line galaxies}
In galaxies of the first group, the \PP EW  is of the order of 0.04-0.25 $\mu m$, its strength increases with increasing color excess, the dust extinction is moderate (average E(B-V) = 0.32) and SFRs range from 0.1 to 10 $M_\odot/yr$. The SFR from \PP is about a factor of three higher than that inferred from \Ha. 
This group most likely includes normal star forming galaxies, where H II regions, molecular gas, and photodissociation regions are spatially well mixed \citep{Imanishi2008}. \cite{Moorwood1986} have shown that the  \PP EW values in starbursts have an average value of 0.1$\mu m$, with some scatter, but never become lower than 0.04$\mu m$. Based on Spitzer/IRS spectra, \cite{Inami2018} proposed a revised starburst/AGN diagnostic diagram, pushing the limit to \PP EW $>$0.06 for starburst. According to this new definition, 22383 and 10512 might host an AGN. Both galaxies are close to the red sequence separation and do not show any particular behavior, so we can conclude that the evidence for AGN is not particularly compelling.
The detection of the \PP suggests the presence of the neutral, smallest PAH population in the interstellar medium (ISM), typically with a radius of $\sim$5 \AA{} or N$_C$ $\sim$ 50 carbon atoms \citep{Schutte1993, Draine2007}.

We investigate if the measured values of the \PP EW are sufficient to justify by themselves the observed color excess. We can estimate the maximum contribution of the \PP EW to the F444W flux, given the measured observed values. We  assume that $F_{F444W} = F_{{PAH_{3.3}}, cont} \times \Delta F_{F444W} + F_{PAH_{3.3, obs}}$, with $F_{F444W}$ measured NIRCam flux in the F444W filter, $F_{{PAH_{3.3}}, cont}$ observed continuum flux measured on the NIRSpec spectrum at the \PP wavelength,  $\Delta F_{F444W}$ width of the  NIRCAM F444W filter and $F_{PAH_{3.3, obs}}$ observed \PP measured flux. 
Considering that we can express the \PP flux as \PP EW$\times F_{{PAH_{3.3}}, cont}$, for a \PP EW = 0.1$\mu m$ at the cluster redshift z=0.3068, corresponding to an observed value of 0.1307$\mu m$,  and $\Delta F_{F444W}$= 1.024 $\mu m$, the contribution of the line to the total F444W flux is 12.7\%, which translates into a variation in galaxy color of 0.13 mag. This simple calculation indicates that while the PAH line is certainly non negligible, its presence alone can not explain the color excess, which can be as high as 0.6 mag. A contribution from the dust component is hence needed to explain the colors.

\subsubsection{Weak emission line galaxies}
\begin{figure}
    \centering
    \includegraphics[width=0.85\linewidth]{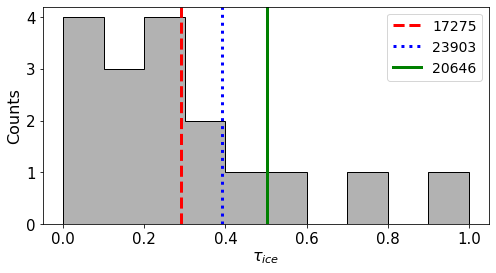}
    \caption{Distribution of the water absorption feature $\tau_{ice}$ for the emission line and weak emission line galaxies, as measured from the CAFE fit. The values of the three weak emission line galaxies are highlighted.
    \label{fig:tau_ice}}
\end{figure}
As mentioned before, 17275, 23903 and 20646 are the three red excess galaxies that, despite having \PP  in emission, have rather weak other emission lines  (S/N of the H$\alpha$ line between 5 and 10). The same three galaxies are the only ones above the red sequence in Fig.\ref{fig:cmd} and two of them are also outliers in the F200W-F444W vs \PP EW plane (Fig.\ref{fig:color_EW}). One of these galaxies host an AGN, based on the MUSE spectra \citep{Vulcani2023a}, while no optical spectra are available for the other two galaxies. Even though their \PP EW is $>$0.06 $\mu m$, our results might  suggest they are all obscured AGN-starburst composite galaxies. Indeed, 
if a highly dust-obscured AGN is present, then absorption features at 3.1 and 3.4 $\mu m$ produced by ice-covered dust and bare carbonaceous dust, respectively, should be found \citep{Imanishi2000,  Imanishi2003, Risaliti2003,Risaliti2006, Imanishi2006b}. According to the CAFE fit, Fig.\ref{fig:tau_ice} shows that these three weak emission line galaxies have rather strong ice absorption, even though are not the strongest. We can therefore not firmly establish the presence of AGN.  It is interesting to mention that the galaxy  wit the strongest absorption feature is 6436, the galaxy with the strongest $A_V$ value that has been also classified as ram pressure stripped galaxy.

\subsubsection{Passive galaxies}\label{sec:pass}
Eleven galaxies in our sample have no significant emission lines (S/N of the H$\alpha$ line $<5$). This group includes all the eight galaxies that are  not red outliers. These non red excess galaxies were selected for lying in various regions of the color-mass and UVJ diagram, but also the galaxies that had blue colors or were supposed to be star forming, turned out to be passive.   We remind the reader, though, that this sample is not representative of all galaxies on the color-mass diagram. As explained in sec. \ref{sec:data}, our targets were used as fillers in different programs, hence they did not have high priority in the preparation of the NIRSpec/MSA masks. As a result, we have no cases in our sample  of star forming galaxies with blue rest frame optical colors and normal F200W-F444W color. From our sample, we can  conclude that galaxies with normal F200W-F444W color have no strong \PP emission.

On the other hand, we found three galaxies (11984, 22207, 5149) that despite having red F200W-F444W color, do not have \PP, nor other lines, in emission. While it is true that these galaxies lie close to the adopted separation between normal and red outliers {  and could be simply misclassified}, other galaxies with similar red excess do show \PP in emission. 
In these cases, red colors are simply related to the red colors of the evolved stellar populations. {  Indeed, they are consistent with the colors expected  from pure stellar models. We calculated the observed F200W-F444W color for a single stellar population (SSP) model at various metallicities (Charlot and Bruzual, priv. comm.) and found that colors above -0.8 can be easily achieved, for both relatively low metallicity, and even for solar values (Z=0.017).

We also  exploited the spectral synthesis code SImulatiNg OPtical Spectra wIth Stellar populations models (SINOPSIS; \citealt{Fritz2007, Fritz2011, Fritz2017}) to look for differences in the stellar age properties  between these three galaxies and the other normal red sequence galaxies. We found no clear differences in terms of ages or star formation histories between the two groups. Given the small sample sizes and the resolution of the spectra, it is hard to establish if this lack is real or due to statistics or data quality and we refrain from drawing solid conclusions.}

\subsection{The discrepancy between H$\alpha$ and \PP based SFRs}\label{sec:disc_SFR}

In Sec. \ref{sec:SFR} we have found  a clear correlation  between  the SFRs obtained from H$\alpha$, corrected for dust based on the Hydrogen lines ratio,  and those obtained from the \PP. Nonetheless, the values obtained from the \PP are systematically larger, by a factor of 3.5, on average.

One possible explanation is that the dust correction is 
insufficient, hence H$\alpha$ based values are underestimated. This can happen if the intrinsic Paschen/Balmer ratio is not sufficient to capture all the dust enshrouding the star formation, or that the geometry of the dust is different at the H$\alpha$ and \PP wavelengths.

\begin{figure}
    \centering
    \includegraphics[width=0.9\linewidth]{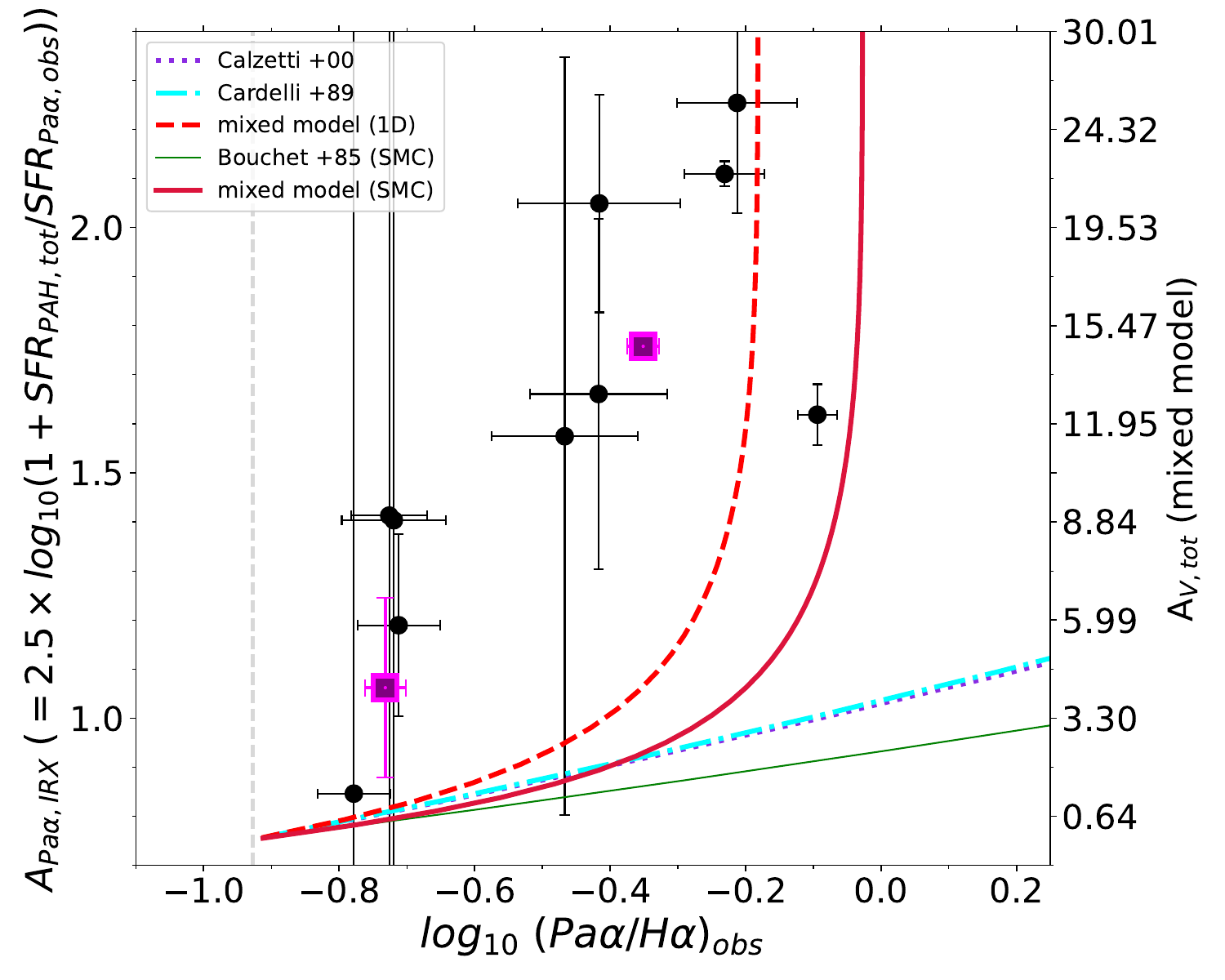}
    \caption{Diagram comparing the ratio of H$\alpha$ and Pa$\alpha$  and A$_{Pa\alpha,IRX}$, that is the ratio of SFRs derived from the observed Pa$\alpha$ and the PAH, (A$_{Pa\alpha,IRX}$=2.5$\times \log(1+SFR_{PAH, tot}/SFR_{Pa\alpha,obs}$). Purple squares represent weighted average values in bins of  Pa$\alpha$/H$\alpha$ ratios. Expected values obtained considering different attenuation laws are also overplotted, as explained in the label.  }
    \label{fig:SFR_check}
\end{figure}

As largely discussed in \cite{Lai2020}, different  geometries can be explored in the modeling for the MIR spectral decomposition. Two reliable examples are the ``fully mixed'' and ``obscured continuum'' geometries, which should  bracket realistic scenarios of PAH emission and silicate absorption in galaxies. Briefly, in the fully mixed case stars and dust are spatially well mixed, with both PAH+line and continuum emission subject to a similar range of attenuation. In contrast, the obscured continuum scheme consists of a concentrated continuum emitting source plus unobscured PAH emitters atop the continuum. In such a case, PAH band strengths are measured with their local continua, and their emission sources are presumed to lie in the “foreground” of the extinction-impact continuum sources. Thus, PAH emission remain free from extinction correction. The former case typically holds for normal star forming regions, while the latter  is typically invoked when galaxies are powered by a buried AGN or compact central starburst with only H II regions, because  in these cases the obscuration comes predominantly from the central part of the system, while the PAH/line emissions excited by stars in the outer part are mostly not attenuated \citep{Imanishi2007, Imanishi2008}. 
To identify the most probable scenario, following \cite{Calabro2018}, in Fig.\ref{fig:SFR_check} we consider only the galaxies with reliable  H$\alpha$ and Pa$\alpha$ measurements and compare  the ratio of H$\alpha$ and Pa$\alpha$ (Paschen-Balmer decrement) to the ratio of SFRs derived from the observed Pa$\alpha$ and the PAH, (A$_{Pa\alpha,IRX}$=2.5$\times \log(1+SFR_{PAH, tot}/SFR_{Pa\alpha,obs}$), assuming it traces the total SFR. These two ratios, both independent measures of attenuation, do not generally scale as predicted by the \cite{Calzetti2000} and \cite{Cardelli1989} attenuation curves, suggesting those models are not a good representation of the dust distribution in these galaxies. The value of Pa$\alpha$/H$\alpha$ rather saturates around -0.1, qualitatively consistent with an optically thick ``mixed model'' in which different lines probe different optical depths. A more thorough analysis is beyond the scope of this paper and will be addressed in a forthcoming study. 

Alternatively, there could be a bias in the star formation traced by the \PP.
PAH emission can indeed be used as star formation tracer in galaxies only  as long as the excitation of PAH molecules is mostly due to UV photons from the massive young stars \citep[e.g.][]{Peeters2004, XieHo2019}. However, the nature and the properties of the PAH molecules remain poorly understood and this could affect the use of PAH emission as an indicator of star formation activity. 

Studies have explored the effect of star formation on the properties of the associated PAH molecules. For example, \cite{Riener2018} investigated the impact of high-energy photons on the dust particles associated with HII regions that are linked to star formation. \cite{Chastenet2023} used JWST observations to explore the properties of PAHs associated with HII regions, finding evidence for heating and/or changes in PAH size in regions with higher molecular gas content as well as increased ionization in regions with higher H$\alpha$ intensity.

\cite{Ujjwal2024} combined JWST and UVIT observations and analyzed how the star formation influences the properties of PAH molecules that are spatially associated with the regions undergoing active star formation. They concluded that the PAH molecules excited by the photons from star-forming regions with higher SFR surface density are significantly smaller and ionized molecules. UV photons from the star-forming regions could be the reason for the higher fraction of the ionized PAHs. They suggested that the effect of the high temperature combined to the formation of smaller PAH molecules in the star-forming regions might also result in the higher emission at 3.3$\mu m$.

Finally, JWST observations also showed diffuse extended PAH emission which is not spatially associated with the star forming regions, suggesting that the PAH can be heated by interstellar radiation field \citep[e.g.][]{Evans2022}. Therefore the observed total \PP may also include the contribution from interstellar radiation field, which, if not correctly taken into account, would bias the SFR estimates.

\section{Summary} \label{sec:summary}
\cite{Vulcani2023a} have unveiled the existence of a population of galaxies with surprisingly red F200W-F444W color that does not stands out in any other integrated property. This population exists in both the cluster A2744 and surrounding field.  In this paper we further characterized this population, thanks to new spectroscopic observations taken with JWST/NIRSpec. The main results can be summarized as follows:
\begin{itemize}
    \item  Red outliers -- defined as galaxies whose F200W-F444W color is redder than 3$\times$ the width of the red sequence of the entire population -- can be subdivided in three groups: i) emission line galaxies with also a clear \PP in emission, which are mostly located below (i.e. bluer than) the optical red sequence; ii)  galaxies with \PP in emission, but rather weak emission lines (S/N of the H$\alpha$ line between 5 and 10), mostly located above (i.e. redder than) the optical red sequence; iii) passive galaxies. The first group is the most common.  
    \item In the emission line galaxies, the F200W-F444W color excess correlates with the EW of the \PP: the higher the excess, the higher the strength of the EW. Nonetheless, the presence of the PAH alone  can not fully explain the color excess, as an EW of $\sim0.1\mu m$ is expected to increase the color by only 0.13 mag.
    \item The weak emission line galaxies are also outliers in the F200W-F444W vs EW(\PP) plane and are located above the red sequence. Spectral fitting identify a strong ice absorption features. This pieces of evidence point to the presence of an AGN. 
    \item The passive galaxies have no \PP. Their red color is due to ageing of the stellar population. 
    \item The non red outlier galaxies in our sample are all passive and have no \PP in emission, confirming that the \PP feature is linked to star formation activity.
    \item Even within our small sample, a variety of \PP profiles emerges, suggesting a variety of emitting molecules with a range of sizes. Also the relative importance of the 3.4 and 3.45 $\mu m$  features and the 3.3 $\mu m$ aromatic feature changes from galaxy to galaxy, indicative of a different chemical composition of the dust particles. 
    \item In the star forming galaxies, both the intrinsic \PP EW and flux correlate with the intrinsic H$\alpha$ EW and flux, suggesting that they are produced by the same mechanism. Nonetheless, the dust corrected SFR from \PP is a factor of 3 higher than the SFR obtained from H$\alpha$, suggesting red outliers are characterized by a significant amount of dust, that is not captured by the ratio of hydrogen lines.
    \item {  Overall, no clear differences are detected between cluster and field galaxies, in terms of occurrence and properties of the \PP emission lines and of the galaxies. This result, though, is  most likely due to the small size of our sample and the way targets were selected, so larger sample will be needed to explore further this aspect. }
\end{itemize}

Our analysis unveils that what we called red outliers is actually a mixed population of normal dusty star forming galaxies, AGN and passive systems. Combining the color selection in the NIR, as done in Fig.\ref{fig:cmd}, and more traditional color-mass and UVJ diagrams can provide a  cleaner sample of star forming systems.

\begin{acknowledgements}
We thank the referee for their useful suggestions that helped us improving the manuscript. We thank Guido Roberts-Borsani for his help on data reduction and calibration and Sedona Price and the UNCOVER team for sharing their spectra with us.
B.V. thanks the Department of Physics and Astronomy, University of California, Los Angeles, for a very pleasant and productive stay during which most of the work presented in this paper was carried out. The data were obtained from the Mikulski Archive for Space Telescopes at the Space Telescope Science Institute, which is operated by the Association of Universities for Research in Astronomy, Inc., under NASA contract NAS 5-03127 for JWST. These observations are associated with programs JWST-ERS-1324, JWST-GO-3073, and JWST-GO-2561. The specific observations associated with program JWST-GO-3073  can be accessed via \href{https://doi.org/10.17909/4r6b-bx96}{DOI}. We acknowledge financial support from NASA through grants JWST-ERS-1324 and JWST-GO-3073.  This project has received funding from the European Research Council (ERC) under the Horizon 2020 research and innovation programme (grant agreement N. 833824). We also  acknowledge support from the INAF Large Grant 2022 “Extragalactic Surveys with JWST” (PI Pentericci), from PRIN 2022 MUR project 2022CB3PJ3 – First Light And Galaxy aSsembly (FLAGS) funded by the European Union – Next Generation EU, and from INAF Mini-grant ``Reionization and Fundamental Cosmology with High-Redshift Galaxies". B.V. is supported  by the European Union – NextGenerationEU RFF M4C2 1.1 PRIN 2022 project 2022ZSL4BL INSIGHT. PS acknowledges INAF Mini Grant 2022 “The evolution of passive galaxies through cosmic time”.

\end{acknowledgements}

\bibliography{references.bib}{}
\bibliographystyle{aa}

\begin{appendix}\label{app:add}
\section{Additional Table and Figure}

Table \ref{tab:galaxies} presents the sample and reports the extinction corrected measurements of the most important lines.

\begin{table*}
    \centering
    \small
\begin{tabular}{r|rrrrrrrr}
\hline
  \multicolumn{1}{c|}{ID} &
  \multicolumn{1}{c}{RA2000} &
  \multicolumn{1}{c}{DEC2000} &
  \multicolumn{1}{c}{specz} &
  \multicolumn{1}{c}{env} &
  \multicolumn{1}{c}{$f_{H\alpha}$ [$10^{-16}$]} &
  \multicolumn{1}{c}{$f_{Pa\alpha}$ [$10^{-16}$]} &
  \multicolumn{1}{c}{$f_{PAH 3.3}$ [$10^{-16}$]} &
  \multicolumn{1}{c}{$EW_{PAH 3.3}$}\\
  \multicolumn{1}{c|}{} &
  \multicolumn{1}{c}{[deg]} &
  \multicolumn{1}{c}{[deg]} &
  \multicolumn{1}{c}{} &
  \multicolumn{1}{c}{} &
  \multicolumn{1}{c}{[$erg/cm^2/s$]} &
  \multicolumn{1}{c}{[$erg/cm^2/s$]} &
  \multicolumn{1}{c}{[$erg/cm^2/s$]} &
  \multicolumn{1}{c}{[$\mu m$]}  \\
  \hline
  4634 & 00:14:21.68 & -30:24:01.4 & 0.4971 & fie & 137$\pm$14 & 16$\pm$1 & 83$\pm$4 & 0.0941\\
  5149 & 00:14:18.98 & -30:24:00.3 & 0.3056 & cl & -- & -- & -- & -- \\
  6436 & 00:14:19.43 & -30:23:26.9 & 0.2931 & cl & 96$\pm$4 & 11.3$\pm$0.5 & 36$\pm$0.2 & 0.1252\\
  7423 & 00:14:16.63 & -30:23:03.3 & 0.2961 & cl & 23$\pm$2 & 2.7$\pm$0.3 & 6.2$\pm$0.4 & 0.1359\\
  7622 & 00:14:25.06 & -30:23:05.9 & 0.2958 & cl & -- & 0.192$\pm$0.003 & 0.645$\pm$0.008 & 0.1071\\
  10512 & 00:14:06.07 & -30:22:23.1 & 0.3143 & cl & 0.65$\pm$0.06 & 0.077$\pm$0.006 & 0.10$\pm$0.02 & 0.0385\\
  11580 & 00:14:06.57 & -30:21:59.1 & 0.3072 & cl  & -- & -- & -- & -- \\
  11984 & 00:14:07.04 & -30:21:53.4 & 0.3067 & cl  & -- & -- & -- & -- \\
  13250 & 00:14:07.72 & -30:21:37.4 & 0.3004 & cl  & -- & -- & -- & -- \\
  13453 & 00:14:07.36 & -30:21:38.1 & 0.28 & fie  & -- & -- & -- & -- \\
  14869 & 00:13:57.01 & -30:21:19.3 & 0.2324 & fie & -- & -- & -- & -- \\
  15448 & 00:13:53.28 & -30:21:01.2 & 0.3058 & cl & 14$\pm$1.43 & 1.7$\pm$0.3 &5.7$\pm$0.2 & 0.0482\\
  16695 & 00:13:59.77 & -30:20:46.4 & 0.455 & fie & 28$\pm$1 & -- & 5.9$\pm$0.5 & 0.1466\\
  17019 & 00:14:00.01 & -30:20:45.5 & 0.4509 & fie & 33$\pm$2 & -- & 25$\pm$2 & 0.2153\\
  17275 & 00:13:52.89 & -30:20:36.9 & 0.49957 & fie & 7$\pm$1 & 0.8$\pm$0.1 & 5.3$\pm$0.2 & 0.0687\\
  17429 & 00:14:08.85 & -30:20:30.7 & 0.2824 & fie & -- & -- & -- & --\\
  18066 & 00:14:08.50 & -30:20:33.1 & 0.3143 & cl & -- & -- & -- & --\\
  20646 & 00:14:01.03 & -30:18:31.3 & 0.3058 & cl & 6$\pm$0.8 & 0.7$\pm$0.2 & 2.5$\pm$0.1 & 0.0458\\
  21206 & 00:13:57.27 & -30:19:27.7 & 0.3072 & cl & 2.5$\pm$0.4 & 0.30$\pm$0.03 & 0.9$\pm$0.2 & 0.0634\\
  21319 & 00:13:58.50 & -30:18:31.7 & 0.4536 & fie & 13.1$\pm$0.5 & 0.59$\pm$0.03 & 2.4$\pm$0.4 & 0.082\\
  21382 & 00:13:57.13 & -30:19:13.2 & 0.3075 & cl & 2.3$\pm$0.2 & 0.2$\pm$0.03 & 0.81$\pm$0.06 & 0.0575\\
  21466 & 00:13:58.07 & -30:18:43.6 & 0.4137 & fie & -- & -- & -- & --\\
  22197 & 00:14:00.18 & -30:18:26.0 & 0.307 & cl & -- & -- & -- & --\\
  22207 & 00:14:03.57 & -30:19:10.3 & 0.46 & fie & -- & -- & -- & --\\
  22383 & 00:13:59.42 & -30:19:14.3 & 0.3498 & fie & 3.7$\pm$0.3 & 0.152$\pm$0.007 & 0.43$\pm$0.07 & 0.0385\\
  23128 & 00:13:56.90 & -30:19:29.2 & 0.5414 & fie & 4.2$\pm$0.2 & ---  & 1.6$\pm$0.9 & 0.1299\\
  23903 & 00:13:59.29 & -30:19:18.8 & 0.3432 & fie & 18$\pm$2 & 2.1$\pm$0.6 & 11.9$\pm$0.6 & 0.0624\\
  24012 & 00:13:59.32 & -30:19:07.7 & 0.343 & fie & 11.6$\pm$0.7 & 0.6$\pm$0.04 & 1.3$\pm$0.1 & 0.0822\\
\hline
  \\
\hline
  \multicolumn{1}{c|}{ID} &
   \multicolumn{1}{c}{$f_{3.4}$ [$10^{-16}$]} &
  \multicolumn{1}{c}{$EW_{3.4}$} &
  \multicolumn{1}{c}{$f_{3.45}$ [$10^{-16}$]} &
  \multicolumn{1}{c}{$EW_{3.45}$} &
   \multicolumn{1}{c}{$f_{3.4}/f_{3.3}$} &
   \multicolumn{1}{c}{$f_{3.45}/f_{3.3}$} &
   \multicolumn{1}{c}{$\tau_{ice}$} &
  \multicolumn{1}{c}{A$_V$}  \\
  \multicolumn{1}{c|}{} &
\multicolumn{1}{c}{[$erg/cm^2/s$]} &
  \multicolumn{1}{c}{[$\mu m$]} &
  \multicolumn{1}{c}{[$erg/cm^2/s$]} &
  \multicolumn{1}{c}{[$\mu m$]} &
  \multicolumn{1}{c}{} &
  \multicolumn{1}{c}{} &
  \multicolumn{1}{c}{} &
  \multicolumn{1}{c}{[mag]} \\
  \hline
  4634 & 10$\pm$1 & 0.0122 & 44$\pm$2 & 0.0564 & 0.16 & 0.12 & 0.54 & 2.6\\
  5149 & -- & -- & -- & -- & --& -- & -- & --\\
  6436 & 2.9$\pm$0.2 & 0.0096 & 2$\pm$1 & 0.0070 & 1.0 & 0.09 & 0.07 & 3.1\\
  7423 & 1.07$\pm$0.09 & 0.0244 & 1.6$\pm$0.1 & 0.0380 & 0.44 & 0.18 & 0.28 & 0.8\\
  7622 & 0.135$\pm$0.009 & 0.0248 & 0.13$\pm$0.0 & 0.0250 & 0.15 & 0.21 & 0.21 & 1.3*\\
  10512 & 0.010$\pm$0.006 & 0.0042 & 0.01$\pm$0.0 & 0.0043 & 0.275 & 0.1 & 0.1 & 0.6\\
  11580 & -- & -- & -- & -- & --& -- & -- & --\\
  11984 & -- & -- & -- & -- & --& -- & -- & --\\
  13250  & -- & -- & -- & -- & --& -- & -- & --\\
  13453  & -- & -- & -- & -- & --& -- & -- & --\\
  14869  & -- & -- & -- & -- & --& -- & -- & --\\
  15448 & 0.62$\pm$0.03 & 0.0056 & 0.95$\pm$0.04 & 0.0090 & 0.19 & 0.11 & 0.17 & 1.9\\
  16695 & 0.9$\pm$0.9 & 0.0246 & 2$\pm$2 & 0.0700 & 0.0 & 0.16 & 0.46 & 1.3*\\
  17019 & 3.0$\pm$0.2 & 0.0258 & 6$\pm$2 & 0.0493 & 0.77 & 0.14 & 0.27 & 1.3*\\
  17275 & 0.30$\pm$0.01 & 0.0041 & 1$\pm$0.2 & 0.0146 & 0.29 & 0.06 & 0.21 & 2.7\\
  17429 & -- & -- & -- & -- & --& -- & -- & --\\
  18066  & -- & -- & -- & -- & --& -- & -- & --\\
  20646 & 0.27$\pm$0.02 & 0.0052 & 0.28$\pm$0.02 & 0.0054 & 0.50 & 0.12 & 0.12 & 1.7\\
  21206 & 0.12$\pm$0.03 & 0.0095 & 0.15$\pm$0.05 & 0.0121 & 0.25 & 0.14 & 0.17 & 0.8\\
  21319 & 0.32$\pm$0.05 & 0.0127 & 1.2$\pm$0.1 & 0.0491 & 0.0 & 0.14 & 0.51 & 1.3*\\
  21382 & 0.13$\pm$0.02 & 0.0101 & 0.17$\pm$0.01 & 0.0141 & 0.3153 & 0.17 & 0.22 & 0.8\\
  21466  & -- & -- & -- & -- & --& -- & -- & --\\
  22197  & -- & -- & -- & -- & --& -- & -- & --\\
  22207  & -- & -- & -- & -- & --& -- & -- & --\\
  22383 & 0.027$\pm$0.005 & 0.0027 & 0.05$\pm$0.01 & 0.00477 & 0.29 & 0.07 & 0.11 & 1.3*\\
  23128 & 0.3$\pm$0.1 & 0.0242 & --  & -- & 0.0 & 0.16 &  & 1.3*\\
  23903 & 0.82$\pm$0.04 & 0.0044 & 1.50$\pm$0.08 & 0.0084 & 0.39 & 0.07 & 0.14 & 1.9\\
  24012 & 0.16$\pm$0.06 & 0.011 & 0.26$\pm$0.03 & 0.0192 & 0.0 & 0.12 & 0.2 & 1.3*\\
\hline
\end{tabular}
    \caption{Spectral properties of the sample. ID and coordinates are from \cite{Paris2023}. Spectroscopic redshfits are from \citet{Vulcani2023a} and from this work. Env is the environment in which galaxies are located (cl = cluster, fi = field). Dust corrected fluxes and equivalent widths are computed as described in the text. A$_V$ is measured from the H$\alpha$/Pa$\alpha$ decrement and using the \cite{Cardelli1989} law. Asterisks indicate that the reported A$_V$ is a mean value, due to the lack of one of the two lines. }
    \label{tab:galaxies}
\end{table*}

\begin{figure*}
    \centering
    \includegraphics[width=0.9\linewidth]{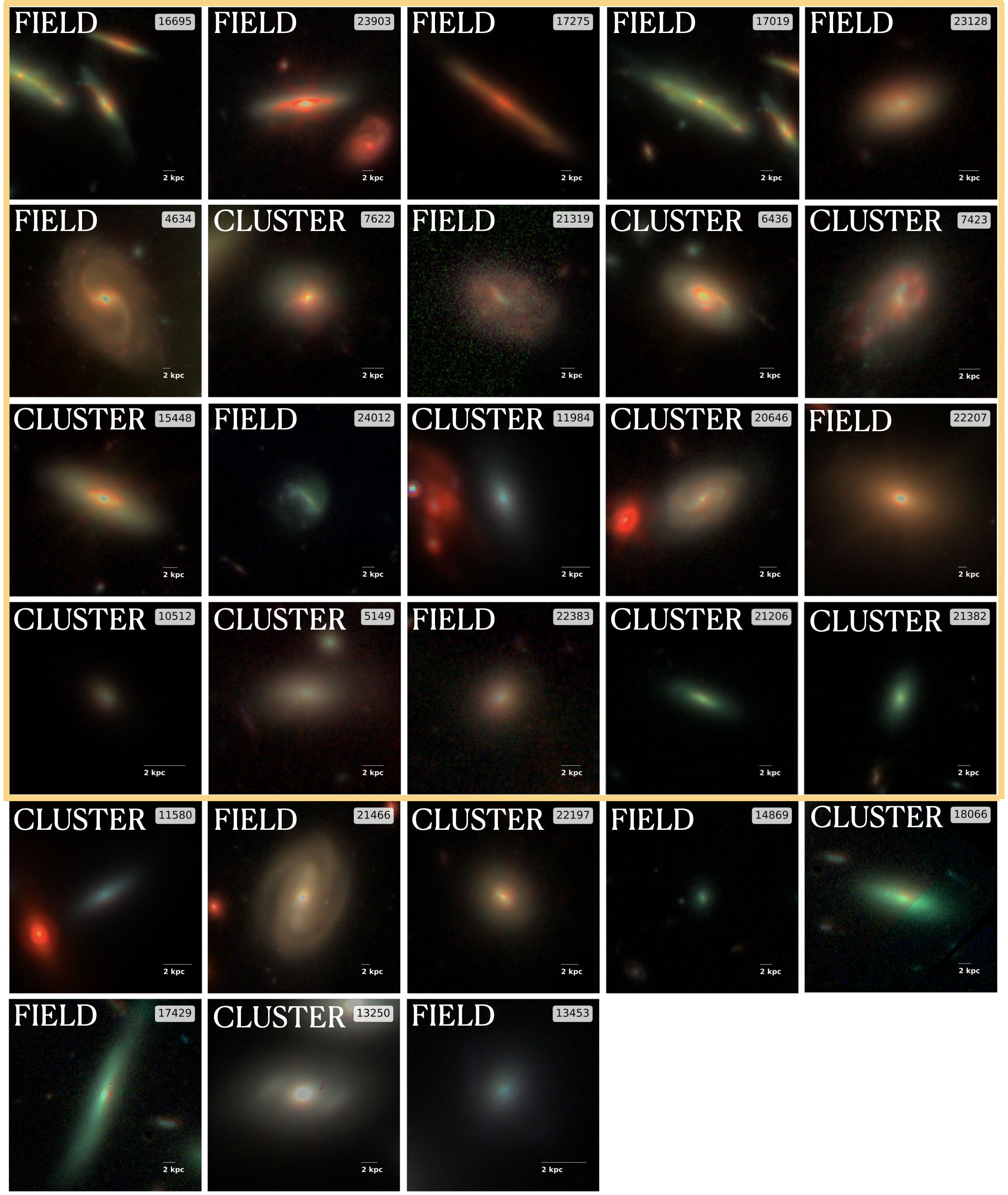}
    \caption{(F115W+F200W (or F150W if F200W is not reliable)+F444W images of the galaxies analyzed in this paper. The physical scale is reported in the lower right corners. Galaxies are sorted by decreasing red excess (see sec. \ref{sec:analysis}). The gold rectangle highlights the red outliers.}
    \label{fig:images}
\end{figure*}

Figure \ref{fig:images} shows  the color-composite images for the cluster and field targets.

\end{appendix}

\end{document}